\documentclass[sigconf]{acmart}

\AtBeginDocument{%
  \providecommand\BibTeX{{%
    \normalfont B\kern-0.5em{\scshape i\kern-0.25em b}\kern-0.8em\TeX}}}
  
\graphicspath{{./figure/}}

\setcopyright{acmcopyright}

\copyrightyear{2019} 
\acmYear{2019} 
\acmConference[BuildSys '19]{The 6th ACM International Conference on Systems for Energy-Efficient Buildings, Cities, and Transportation}{November 13--14, 2019}{New York, NY, USA}
\acmBooktitle{The 6th ACM International Conference on Systems for Energy-Efficient Buildings, Cities, and Transportation (BuildSys '19), November 13--14, 2019, New York, NY, USA}
\acmPrice{15.00}
\acmDOI{10.1145/3360322.3360861}
\acmISBN{978-1-4503-7005-9/19/11}

\usepackage{multirow}
\usepackage{booktabs} 
\usepackage{amsmath}
\usepackage{subcaption}
\usepackage{url}
\usepackage{hyperref}
\usepackage{graphicx}
\usepackage{multirow}
\usepackage[linesnumbered,lined,boxed,commentsnumbered]{algorithm2e}

\DeclareMathOperator*{\argmax}{arg\,max}

\begin{document} 
  

\title[HVAC Scheduling Using RL via NN Based Model Approx.]{Building HVAC Scheduling Using Reinforcement Learning via Neural Network Based Model Approximation}

\titlenote{This work has been sponsored by the U.S. Army Research Office (ARO) under award number W911NF1910362 and the U.S. National Science Foundation (NSF) under award number 1911229.}

\author{Chi Zhang}
\affiliation{%
  \department{Department of Computer Science}
  \institution{University of Southern California}
  \city{Los Angeles}
  \state{CA}
  \postcode{90089}
}
\email{zhan527@usc.edu}

\author{Sanmukh R. Kuppannagari}
\affiliation{%
  \department{Department of Electrical and Computer Engineering}
  \institution{University of Southern California}
  \city{Los Angeles}
  \state{CA}
  \postcode{90089}
}
\email{kuppanna@usc.edu}

\author{Rajgopal Kannan}
\affiliation{%
  \department{US Army Research Lab-West}
  \city{Playa Vista}
  \state{CA}
  \postcode{90094}
}
\email{rajgopak@usc.edu}

\author{Viktor K. Prasanna}
\affiliation{%
  \department{Department of Electrical and Computer Engineering}
  \institution{University of Southern California}
  \city{Los Angeles}
  \state{CA}
  \postcode{90089}
}
\email{prasanna@usc.edu}

\renewcommand{\shortauthors}{Chi Zhang et al.}


\begin{abstract}
  Buildings sector is one of the major consumers of energy in the United States. The buildings HVAC (Heating, Ventilation, and Air Conditioning) systems, whose functionality is to maintain thermal comfort and indoor air quality (IAQ), account for almost half of the energy consumed by the buildings. Thus, intelligent scheduling of the building HVAC system has the potential for tremendous energy and cost savings while ensuring that the control objectives (thermal comfort, air quality) are satisfied.
  
  Traditionally, rule-based and model-based approaches such as linear-quadratic regulator (LQR) have been used for scheduling HVAC. However, the system complexity of HVAC and the dynamism in the building environment limit the accuracy, efficiency and robustness of such methods. Recently, several works have focused on model-free deep reinforcement learning based techniques such as Deep Q-Network (DQN). Such methods require extensive interactions with the environment. Thus, they are impractical to implement in real systems due to low sample efficiency. Safety-aware exploration is another challenge in real systems since certain actions at particular states may result in catastrophic outcomes.
  
  To address these issues and challenges, we propose a model-based reinforcement learning approach that learns the system dynamics using a neural network. Then, we adopt Model Predictive Control (MPC) using the learned system dynamics to perform control with random-sampling shooting method. To ensure safe exploration, we limit the actions within safe range and the maximum absolute change of actions according to prior knowledge. We evaluate our ideas through simulation using widely adopted EnergyPlus tool on a case study consisting of a two zone data-center. Experiments show that the average deviation of the trajectories sampled from the learned dynamics and the ground truth is below $20\%$. Compared with baseline approaches, we reduce the total energy consumption by $17.1\% \sim 21.8\%$. Compared with model-free reinforcement learning approach, we reduce the required number of training steps to converge by 10x.
\end{abstract}





\maketitle

\section{Introduction}
The energy consumption by buildings consist of 40\% of the total energy and 70\% of total electricity in the United States \cite{building_energy_data_book}.
Of the total energy consumption of buildings, the Heating, Ventilation and Air-Conditioning (HVAC) system accounts for 50\% while the rest is used for lighting, electrical appliances, electric vehicles, etc. The main objective of the HVAC system is to maintain the indoor temperature and air quality. An intelligent HVAC scheduling system will, additionally, save energy and cost while satisfying the objective. The HVAC system is a nonlinear system and has complex system dynamics with a large number of subsystems including chillers, boilers, heat pumps, pipes, ducts, fans, pumps and heat exchangers \cite{hvac_intro}.
In this paper, we assume the combination of equipment to operate by the HVAC system is fixed and focus on how to set the temperature point for local controllers to reduce the energy consumption or cost while maintaining the thermal comfort at given level. 

%

Traditional approaches for set point scheduling include PID control \cite{hvac_pid} and rule-based supervisory controllers \cite{supervisory_rule_based}. 
The PID controller is a feedback proportional–integral–derivative controller which works by simply turning on/off the HVAC systems. Some advanced HVAC scheduling systems employ rule-based supervisory controllers given historical operation experience. Such systems require no system modeling and design effort, however, they require an experienced and professional operator to constantly monitor and control the system which increases operational costs. Moreover, these traditional approaches are reactive in nature i.e. they are based on feedback from the system and lack the ability to anticipate how the system evolves. This hinders their energy performance as - (i) the thermal inertia of the building leads to delayed effect of control action requiring more aggressive actions, and (ii) the inability to predict and account for external disturbances such as weather, electricity price and occupancy conditions leads to sub-optimal decision making.

Recent success of deep learning has led to the development of several deep reinforcement learning (DRL) based approaches for HVAC scheduling~\cite{drl_hvac, practical_drl_radiant_heating, rl_gym_env}. These data-driven approaches learn an agent to schedule the HVAC system by interacting with the environment. DRL can be generally divided into model-free approach and model-based approach.
In model-free DRL, the agent learns the policy by directly interacting with the environment. 
The agent explores the environment by extensively trial-and-error. However, for a constrained system such as HVAC which enforces soft constraints of the feasible region of operation i.e. thermal comfort bounds, model-free DRL techniques such as Deep Q-Network~\cite{dqn} require a large amount of operational data (obtained via interactions) to converge (also known as low sample efficiency) which is difficult to gather in a real system.

Thus, the practical alternative is to adopt model-based RL approaches. The model for RL algorithm can be obtained by developing a thermal dynamics model \cite{optimal_control_building_hvac_sequential_quadratic_programming}. However, the complexity of the HVAC system and the dynamism of the building environment makes it a daunting task \cite{building_dynamics_model}. Thus, an alternative is to use the readily available historical data on HVAC operation and learn a general function approximator (e.g. neural network) for building system dynamics~\cite{nn_model_based_rl}. Planning algorithms such as linear-quadratic regulator (LQR) \cite{lqr} can be used on the learned dynamic to perform HVAC scheduling to minimize energy consumption with thermal comfort constraints. 

In this paper, we develop a model-based reinforcement learning approach for smart building HVAC control by learning the system dynamics using operation data to fit a neural network. 
Then, we perform Model Predictive Control (MPC) \cite{mpc} using the learned dynamics with random-sampling shooting method \cite{random_sampling_shooting}.
Compared with model-free approaches, our approach is able to train an accurate model with limited amount of data and achieve good control performance without extensive trial-and-error with the systems.
Compared with manually design model, our approach is more general and applicable to various building models since it learns system dynamics automatically from data. 
Our main contributions are as follows:
\begin{itemize}
	\item We analyze the fundamental drawbacks of previous model-free DRL-based approaches and emphasize the importance of sample data efficiency in data-driven approaches.
	\item We propose a model-based DRL approach for building HVAC control that trains the system dynamics with neural networks online. Our approach is both data efficient and self-adaptive online with gradually changing system dynamics (e.g. outdoor temperature).
	\item Given trained system dynamics, we perform Model Predictive Control (MPC) to produce action for the next step that minimizes the energy cost and the temperature constraints violation collectively with random-sampling shooting method.
	\item To support real-time inference, we train an auxiliary policy network that imitates the output of MPC.
	\item We conduct experiments on a two-room data center and show that our approach reduces the total energy consumption by $17.1\% \sim 21.8\%$. Compared with model-free reinforcement learning approach, we reduce the required number of training steps to converge by 10x.
\end{itemize}

\section{Related Work}
In \cite{building_dynamics_model}, the author proposes to estimate the thermal load and use a regulator and a disturbance rejection component to keep the room at comfort temperature. In \cite{novel_modeling}, the author proposes a novel multi-input multi-output (MIMO) to model HVAC system and uses a linear-quadratic regulator (LQR) \cite{lqr} to optimize control performance and to stabilize the proposed HVAC system. In general, exact modeling of HVAC system dynamics is difficult and several data-driven approaches are proposed recently. 
In \cite{data_center_cooling_mpc}, the authors propose linear models for system identification and limit each control variable to a safe range informed by historical data for exploration. Then, the authors minimize a length-$L$ trajectory to obtain the control sequence, execute the action at first step and re-run the optimization. Although they achieve significant performance improvement over baseline controllers, they did not consider distribution shift over time, which requires data aggregation and model fine-tuning.
In \cite{drl_hvac} and \cite{thermal_comfort_rl}, the author proposes to use Deep Q-Network \cite{dqn} to control HVAC systems. In \cite{rl_gym_env}, the author proposes a EnergyPlus based research environment for developing reinforcement learning approaches for data-center HVAC control. In \cite{appro_mpc_nn}, the author shows promising results of approximating the Model Predictive Controller using neural networks. In \cite{ann_mpc_review}, the author proposes to learn system dynamics using Artificial Neural Networks (ANN) and perform model predictive control \cite{mpc}. The main drawbacks of their approach lies in that they attempt to learn everything offline, which fails to adapt to system distribution shift.
\section{Building HVAC System Modeling}
\begin{figure}
  \centering
  \includegraphics[width=\linewidth]{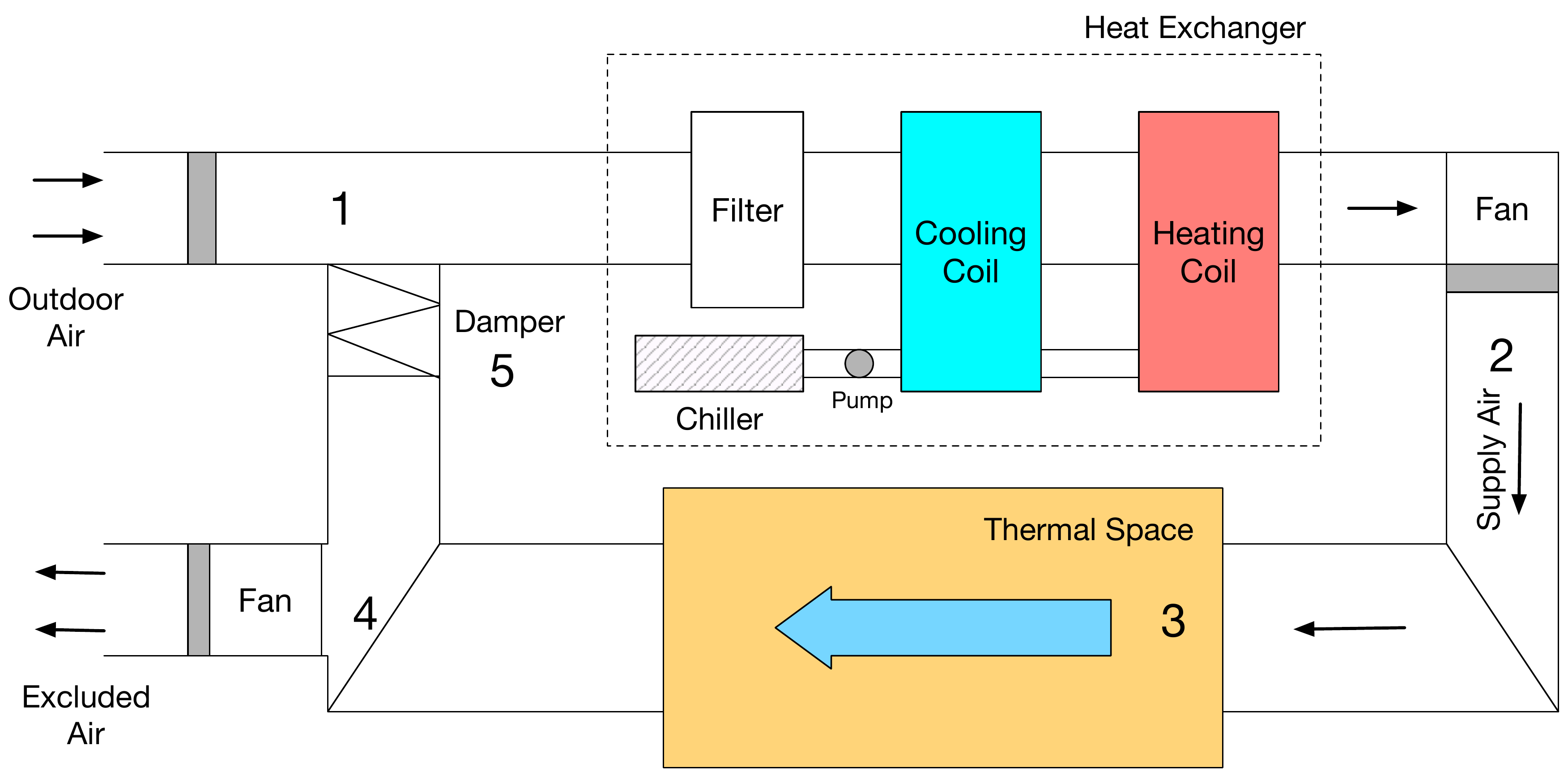}
  \caption{Model of typical single-zone building HVAC system \cite{building_dynamics_model}}
  \label{fig:typical_hvac_system}
\end{figure}

In this section, we demonstrate a typical single-zone building HVAC system that is extensively studied in \cite{building_dynamics_model}. We analyze the system using Control System Equations \cite{control_system_equations} and generalize it using Partial Observable Markov Decision Process \cite{pomdp}. Then, we emphasize the importance and advantages of data-driven approach for this problem and introduce modern reinforcement learning based control strategies.

\subsection{Representative System Modeling}\label{subsection:representation_system}
We show the representative single-zone system model in Figure~\ref{fig:typical_hvac_system}. 
It consists of the following components: a heat exchanger; a chiller, which provides chilled water to the heat exchanger; a circulating air fan; the thermal space; connecting ductwork; dampers; and mixing air components \cite{building_dynamics_model}. The operating mode of the representative system is as follows \cite{building_dynamics_model}:
\begin{itemize}
  \item First, At position 1 as shown in Figure~\ref{fig:typical_hvac_system}, 25\% of the fresh air and 75\% of the recirculated air from position 5 is mixed at the flow mixed.
  \item Second, air mixed at the flow mixer (position 1) enters the heat exchanger, where it is conditioned.
  \item Third, the conditioned air is moved out of the heat exchanger as shown in position 2. This air is ready to enter the thermal space.
  \item Fourth, the supply air enters the thermal space in position 3 and offsets the sensible (actual heat) and latent (humidity) heat loads acting upon the system.
  \item Finally, the air in the thermal space is drawn through a fan as shown in position 4. 75\% of the air is recirculated and 25\% is exhausted to the outdoor environment.
\end{itemize}
The control variables in this system are 1) the speed of the variable-air-volume (VAV) fan as shown in position 2. 2) the water flow rate from the chiller to the heat exchanger. The control system equations can be derived from energy conservation principles and are shown in \cite{building_dynamics_model}:
\begin{align}
\dot{T_3}=&\frac{f}{V_s}(T_2-T_3)-\frac{h_{fg}f}{C_pV_s}(W_s-W_3) + \frac{1}{0.25C_pV_s}(Q_o-h_{fg}M_o)\nonumber\\
\dot{W_3}=&\frac{f}{V_s}(W_s-W_3)+\frac{M_o}{\rho V_s}\nonumber\\
\dot{T_2}=&\frac{f}{V_{he}}(T_3-T_2)+\frac{0.25f}{V_{he}}(T_o-T_3)\nonumber\\
&-\frac{fh_w}{C_pV_{he}}((0.25W_o+0.75W_3)-W_s)-6000\frac{\text{gpm}}{pC_pV_{he}}
\label{eq:manual_dynamics}
\end{align}
where $h_{w}$ is the enthalpy of liquid water, $W_o$ is the humidity ratio of outdoor air, $h_{fg}$ is the enthalpy of water vapor, $V_{he}$ is the volume of heat exchanger, $W_s$ is the humidity ratio of supply air, $W_3$ is the humidity ratio of thermal space, $C_p$ is the specific heat of air, $T_o$ is the temperature of outdoor air, $M_o$ is the moisture load, $Q_o$ is the sensible heat load, $T_2$ is the temperature of supply air, $T_3$ is the temperature of thermal space, $V_s$ is the volume of thermal space, $\rho$ is the air mass density, $f$ is the volumetric flow rate of air (ft /min) and gpm is the flow rate of chilled water (gal/min). Among them, f and gpm are control variables, $T_2$, $T_3$ and $W_3$ are sensible states, $Q_o$ and $M_o$ are latent/hidden states and the rest are system parameters. In \cite{building_dynamics_model}, the author proposes a reduced-order observer as an estimate of the latent/hidden/unmeasurable states. Then, a disturbance rejection controller is proposed to solve the linear time-invariant state feedback system. However, there are several drawbacks of this approach:
\begin{itemize}
  \item The HVAC system is modeled as linear systems due to many perfect systems assumptions \cite{building_dynamics_model}. However, they may not be true in real systems.
  \item The latent states are infeasible to measure at run time.
  \item The mathematical equations are only applicable to this system dynamics and we need to derive manually when it changes. It is even worse that some real system are too complicated to model using equations.
\end{itemize} 
This leads us to the model-based reinforcement learning approach, where we directly learn system transition model using data. Before that, we introduce the Partial Observable Markov Decision Process (POMDP) that reinforcement learning algorithms solve.

\subsection{Partial Observable Markov Decision Process}
\subsubsection{Notations}
We summarize the notations used in this paper as follows:
\begin{itemize}
  \item $a(t)$: action vector at time step $t$, which is the control variable f and gpm mentioned in Section~\ref{subsection:representation_system}.
  \item $s(t)$: state vector at time step $t$, which is $T_o, M_o, Q_o, T_2, T_3, W_3$ in Section~\ref{subsection:representation_system}.
  \item $o(t)$: observation vector at time step $t$, which is $T_2, T_3, W_3$ in Section~\ref{subsection:representation_system}.
  \item $p(s(t+1)|s(t),a(t))$: state transition function, which is Equation~\ref{eq:manual_dynamics} in Section~\ref{subsection:representation_system}.
  \item $r(t)$: reward at time step $t$. It can be defined as $-c(t)$, where $c(t)$ is the cost function.
  \item $c_i(t)$: constraint $i$ at time step $t$ with upper bound $c_{i,max}$ and lower bound $c_{i,min}$. Typical constraints include comfort temperature bound.
  \item $\gamma$: discount factor in reinforcement learning.
\end{itemize}

\subsubsection{Problem Formulation}
We can rewrite Equation~\ref{eq:manual_dynamics} in discrete time domain and generalize it as follows:
\begin{itemize}
  \item $s(t+1)=f_{sys}(s(t), a(t))$
  \item $o(t)=f_{obs}(s(t))$
  \item $r(t)=f_{out}(o(t), a(t))$
  \item $c_i(t)=f_{cons_i}(o(t), a(t))$
\end{itemize}
where $f_{sys}, f_{obs}, f_{out}, f_{cons_i}$ stands for system dynamics, observation emission, reward/cost function, constraint function $i$, respectively. The objective of POMDP is to maximize discounted reward while satisfying each constraint at each time step:
\begin{equation}
\max\sum_{t=0}^{\infty}\gamma^tr(t), \text{, s.t. } c_{i,min}\leq c_i(t)\leq c_{i,max}, \forall i=1, 2, \cdots, N
\label{eq:objective}
\end{equation}
In most building HVAC systems, the cost function is energy consumption or energy cost and constraints are temperature and humidity range.
By collecting data from actuators and sensors in building HVAC system, we can fit these functions using general function approximators (e.g. neural networks) without knowing exact complex underlying physics and solve constrained trajectory optimization problem \cite{trajectory_optimization}.

Note that the state $s(t)$ is not fully measurable in real systems and $f_{sys}$ cannot be directly fitted. Instead, we assume the observation satisfies $W$-step Markov property and predict the next step observation conditioned on a sliding window of previous $W$ steps observation and actions:
\begin{equation}
\begin{split}
\hat{o}(t+1)=f_{obs,sys}(o(t-W+1:t), a(t-W+1:t))
\end{split}
\label{eq:model_dynamics}
\end{equation}
\section{Reinforcement Learning for Building HVAC Control}
\subsection{Model-Free Approach}
In model-free reinforcement learning (MFRL), the agent interacts with the building environment and optimize the policy directly. 
Since MFRL cannot deal with constraints, reward shaping \cite{reward_shaping} is required to combine both cost and constraints into a single reward signal through penalty. Following \cite{rl_gym_env}, we define our reward function as follows
\begin{equation}
r=r_T+\lambda_P r_P
\label{eq:general_reward_shaping}
\end{equation}
where $r_T$ represents the cost, $r_P$ represents the constraints and $\lambda$ controls the tradeoff between the cost and the constraints. A careful fine-tuning of $\lambda$ is required based on different problem emphasis.
Data efficiency is the major problem in MFRL since it is generally not possible to sample large amount of data in real systems, which makes the agent convergence extremely slow.

In this paper, we train a HVAC controller in our simulated environment using Proximal Policy Optimization \cite{ppo} with modified reward defined in Equation~\ref{eq:general_reward_shaping} as baseline approaches for performance comparison.

\subsection{Model-Based Approach}
\begin{figure*}
  \centering
  \includegraphics[width=\linewidth]{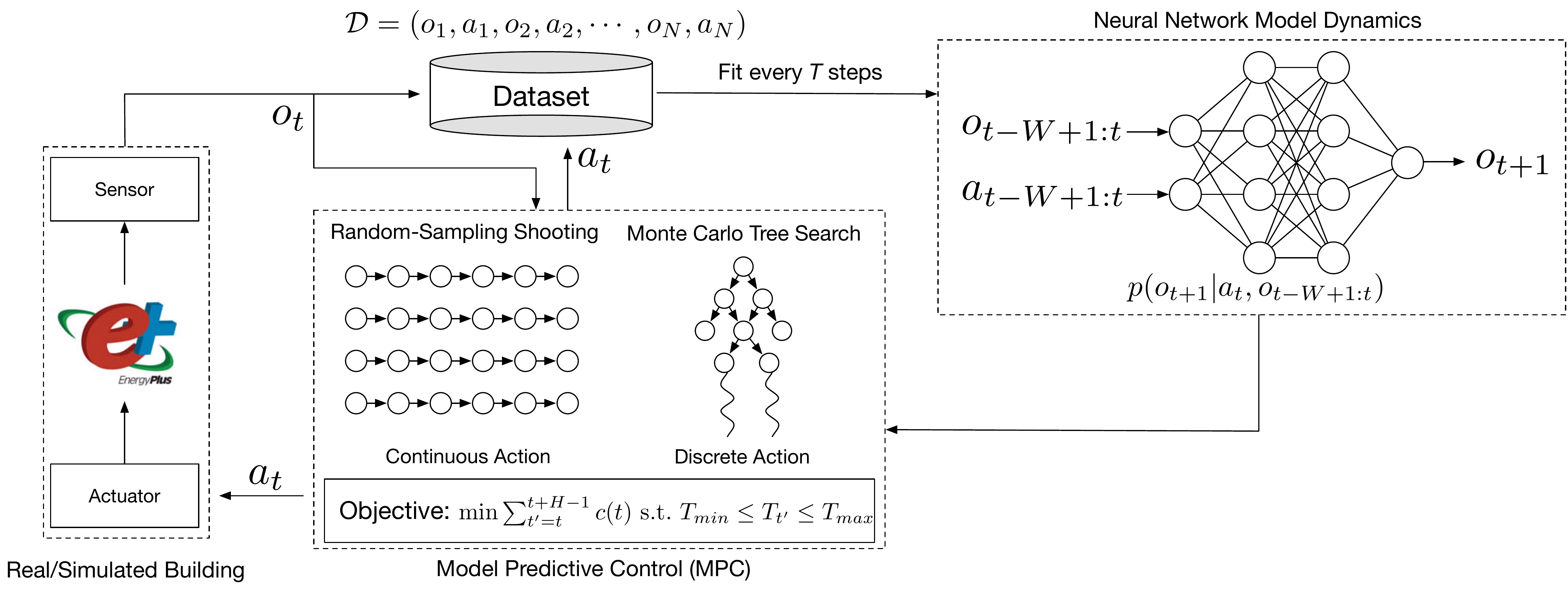}
  \caption{System overview of our model-based reinforcement learning for building HVAC control}
  \label{fig:system_overview}
\end{figure*}

In model-based reinforcement learning (MBRL), the agents learn a system dynamics function by interacting with the systems and use the learned system dynamics to perform trajectory optimization to obtain optimal action sequence. In this section, we elaborate our MBRL approach for building HVAC control.

\subsubsection{System Description}
We illustrate our overall system diagram in Figure~\ref{fig:system_overview} and detailed procedure in Algorithm~\ref{algo:model_rl_hvac_control}. The MBRL agent consists of four parts: dataset $\mathcal{D}$, neural network dynamics model, model predictive control (MPC) and neural network based imitation policy.

\paragraph{\textbf{Model Dynamics Representation}}
We parameterize our model dynamics by $f_{obs,sys}(\cdot;\theta)$ as shown in Equation~\ref{eq:model_dynamics} by a deep neural network, where $\theta$ represents the weights. Following \cite{nn_model_based_rl}, we consider learning deterministic dynamics model by fitting $\Delta o=o(t+1)-o(t)$ instead of $o(t+1)$ since this function approximator would be hard to learn if the adjacent observations are similar and the effect of actions is neglected \cite{nn_model_based_rl}. Advanced stochastic dynamics models including Gaussian Process \cite{gaussian_process} are candidates for future work.

\paragraph{\textbf{Data collection}}
We collect the initial training dataset by executing the default controller (rule-based supervisory or PID) action $a(t)$ and obtain the next observation $o(t+1)$. The dataset $\mathcal{D}$ is a trajectory (execution sequence) of $(o(0),a(0),o(1),a(1),\cdots,o(N-1),a(N-1),o(N))$. Note that this is only the dataset for training the initial dynamics model. Later on, we will add on-policy execution data into $\mathcal{D}$ such that the learned dynamics model can adapt to the potential missing or changes of dynamics distribution.

\paragraph{\textbf{Data preprocessing}}
The neural network based dynamics model takes previous $W$ time step observations $o(t-W+1:t)$ and actions $a(t-W+1:t)$ and output the next observation $o(t+1)$ as shown in Figure~\ref{fig:system_overview}. In building HVAC control, observations can be temperature, humidity ratio, power, etc. These measurements have various range and the weights of the losses will be different if we feed the raw value directly to train the neural network model. Thus, we subtract the mean of the observation/action and divide by the standard deviation as shown in Equation~\ref{eq:normalization}
\begin{equation}
x'=\frac{x-\overline{x}}{\sigma(x)}
\label{eq:normalization}
\end{equation}
where $x$ stands for observation or action.

\paragraph{\textbf{Training dynamics model}} We train the dynamics model by minimizing Mean Square Error (MSE) between predicted delta observation and ground truth delta observation as follows:
\begin{equation}
\begin{split}
\mathcal{E}(\theta)=&\frac{1}{|\mathcal{D}|}\sum_{\substack{o(t-W+1:t)\in\mathcal{D}\\a(t-W+1:t)\in\mathcal{D}}}\frac{1}{2}||o(t+1)\\ &-f_{obs,sys}(o(t-W+1:t),a(t-W+1:t);\theta)||^2
\end{split}
\label{eq:training_dynamics}
\end{equation}
We perform stochastic gradient descent \cite{sgd} on Equation~\ref{eq:training_dynamics} for $M$ epochs. Selecting $M$ is a little bit tricky. Large $M$ may cause $f_{obs,sys}$ to overfit to the current distribution and fail to adapt when new data appended to the dataset. Small $M$ may cause $f_{obs,sys}$ to underfit and deteriorate the performance of Model Predictive Control.

\paragraph{\textbf{Model Predictive Control}}
With learned dynamics model, we perform constrained trajectory optimization of horizon $H$ as follows:
\begin{equation}
\begin{split}
&A_{t}^{H}=\argmax_{A_{t}^{H}}\sum_{t'=t}^{t+H-1} r(\hat{o}(t'), a(t')) :\hat{o}(t)=o(t), \\
&\hat{o}(t'+1)=f_{obs,sys}(o(t'-W+1:t'),a(t'-W+1:t'))
\end{split}
\label{eq:mpc_objective}
\end{equation}

In model predictive control (MPC) \cite{mpc}, we only take the first action Equation~\ref{eq:mpc_objective} returns and rerun constrained trajectory optimization for the next time step.
We consider random-sampling shooting method \cite{random_sampling_shooting} to perform MPC, in which $K$ random action sequences are generated and evaluated by the objective and the constraints using the learned dynamics. Then, we select the one with maximum reward that satisfy the constraints or with minimum violations.


\paragraph{\textbf{Imitating MPC using Neural Networks}}
The main issue of the MPC algorithm is the poor runtime performance and it is infeasible to perform real-time control. Thus, we adopt the idea of DAGGER \cite{dagger} by training a neural network $f_{imit}(\cdot;\phi)$ that clones the output of model predictive controller with on-policy data aggregation. To achieve this, we need another dataset that stores the observation and action pair returned from MPC \textbf{offline}. Then, we minimize the following objective by stochastic gradient descent \cite{sgd}:
\begin{equation}
\min_{\phi}\frac{1}{2}||f_{imit}(o(t-W+1:t);\phi)-a(t)||^2
\label{eq:imitation_objective}
\end{equation}
The detailed procedure is shown in Algorithm~\ref{algo:model_rl_hvac_control_imitation}.
A neural network based policy has extremely fast inference speed for real-time control. However, it may increase constraints violation rate, which is not ideal if it affects system security.

\begin{algorithm}[t]
  \SetAlgoLined
  Gather dataset $\mathcal{D}$ using default policy\;
  Randomly initialize model parameter $f_{obs,sys}(\cdot;\theta)$\;
  \While{True}{
    Fit $f_{obs,sys}(\cdot;\theta)$ using $\mathcal{D}$ on Equation~\ref{eq:training_dynamics} by performing $M$ epochs stochastic gradient descent\;
    \For{i = t : t+T}{
      Obtain building observation $o(t)$ from sensors\;
      Obtain historical observations $o(t-W+1:t-1)$ from $\mathcal{D}$\;
      Solve optimization problem defined in Equation~\ref{eq:mpc_objective} and obtain action sequence $A_{t}^{H}$\;
      Execute the first action $a(t)$ returned from $A_{t}^{H}$\;
      Append ($o(t), a(t)$) to $\mathcal{D}$;
    }
  }
  \caption{Model-based Reinforcement Learning for HVAC Control}
  \label{algo:model_rl_hvac_control}
\end{algorithm}

\begin{algorithm}[t]
  \SetAlgoLined
  Gather dataset $\mathcal{D}$ using default policy\;
  Initialize empty dataset $\mathcal{D'}$ for observation-action pair\;
  Randomly initialize model parameter $f_{obs,sys}(\cdot;\theta)$\;
  Randomly initialize policy parameter $f_{imit}(\cdot;\phi)$\;
  \While{True}{
    Fit $f_{obs,sys}(\cdot;\theta)$ using $\mathcal{D}$ on Equation~\ref{eq:training_dynamics} by performing stochastic gradient descent $M$ epochs\;
    Fit $f_{imit}(\cdot;\phi)$ using $\mathcal{D'}$ on Equation~\ref{eq:imitation_objective} by performing stochastic gradient descent $M$ epochs\;
    \For{i = t : t+T}{
      Obtain building observation $o(t)$ from sensors\;
      Obtain historical observations $o(t-W+1:t-1)$ from $\mathcal{D}$\;
      Compute $a(t)=f_{imit}(o(t-W+1:t);\phi)$ and execute $a(t)$\;
      Append ($o(t), a(t)$) to $\mathcal{D}$\;
      Solve optimization problem defined in Equation~\ref{eq:mpc_objective} and obtain action sequence $A_{t}^{H}$ \textbf{offline}\;
      Append the first action and observation pair $(o(t-W+1:t), a_{mpc}(t))$ to $\mathcal{D'}$;
    }
  }
  \caption{Model-based Reinforcement Learning for HVAC Control with Neural Network Policy}
  \label{algo:model_rl_hvac_control_imitation}
\end{algorithm}

\subsubsection{Discussions and Notes}
\begin{itemize}
  \item \textbf{Dataset replacement policy}: We simply apply First-In-First-Out (FIFO) policy to replace old data. The reason is that $f_{obs,sys}$ needs to adapt to the new system dynamics distribution as the system progresses. In this sense, our approach is \textbf{online learning} and is advantageous to model-free approach.
  \item \textbf{Safety-aware exploration}: Exploration plays a crucial role in reinforcement learning. In model-free reinforcement learning, the agents try novel actions and obtain rewards signal from these actions in order to update policy network. However, certain actions are forbidden in real systems due to security. In this case, a model is necessary for the agent to foresee the outcome.
\end{itemize}
\section{Case Study}
As a case study\footnote{Our implementation is open source at https://github.com/vermouth1992/mbrl-hvac}, we evaluate our model-based reinforcement learning approach on a two-room data center proposed in \cite{rl_gym_env}. The testbed is based on OpenAI Gym \cite{openai_gym} and EnergyPlus \cite{energyplus} and open sourced at \url{https://github.com/IBM/rl-testbed-for-energyplus}.
\subsection{System Modeling}
\subsubsection{Overall Description}
\begin{figure}
  \centering
  \includegraphics[width=\linewidth]{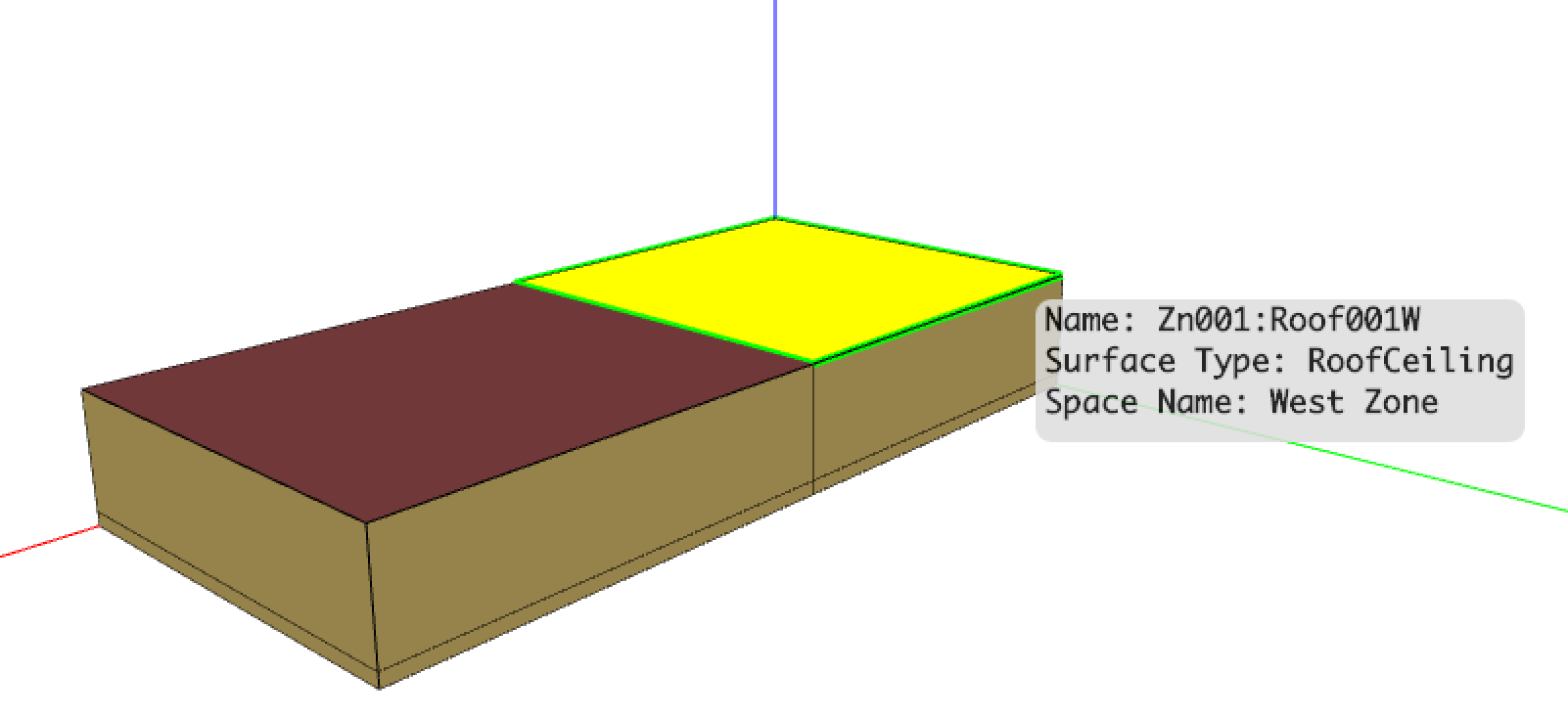}
  \caption{Case study: two room data center}
  \label{fig:two_room_datacenter}
\end{figure}

The target system contains two zones (east zone and west zone), where the thermal load is IT Equipment (ITE) such as servers as shown in Figure~\ref{fig:two_room_datacenter}. Each zone has a dedicated HVAC system similar to Figure~\ref{fig:typical_hvac_system} with the following components: outdoor air system (OA System), variable volume fan (VAV Fan), direct evaporative cooler (DEC), indirect evaporative cooler (IEC), direct expansion cooling coil (DX CC) and chilled water cooling coil (CW CC) \cite{rl_gym_env}. For each zone, the temperature for all the components are specified by a common setpoint. The air volume supplied to each zone is also adjusted by the VAV Fan. 

\subsubsection{POMDP Formulation}\label{sec:pomdp_experiment}
\paragraph{\textbf{Observations}} The observation vector contains:
\begin{itemize}
  \item $T_{out}$: outdoor air temperature
  \item $T_{west}$: west zone air temperature
  \item $T_{east}$: east zone air temperature
  \item $P_{ite}$: IT equipment (ITE) electric demand power
  \item $P_{hvac}$: HVAC electric demand power
\end{itemize}

\paragraph{\textbf{Raw Actions}} The action vector contains:
\begin{itemize}
  \item $TS_{west}$: west zone setpoint temperature
  \item $TS_{east}$: east zone setpoint temperature
  \item $F_{west}$: west Zone supply fan air mass flow rate
  \item $F_{east}$: east Zone supply fan air mass flow rate
\end{itemize}

\paragraph{\textbf{Safety-Aware Exploration and Control Strategy}}
According to our prior knowledge of the building HVAC system, a good temperature setpoint scheduling should fluctuate around the target temperature within safety bounds. Moreover, the maximum absolute changes in setpoint temperature and supply fan air mass flow rate must be limited to ensure hardware security. Following \cite{data_center_cooling_mpc}, we redefine our action space as:
\begin{equation}
a(t)=\text{clip}(\Delta\times z(t)+a(t-1), a_{min}, a_{max})
\end{equation}
where $a(t)$, $a(t-1)$ is the raw action at $t$ and $t-1$. $a_{min}$ and $a_{max}$ are safe action bounds. $\Delta$ is the maximum action change. $z(t)$ is normalized action within $[-1, 1]$.

\paragraph{\textbf{Objective}} Minimize the total power consumption $P_{total}$, where $P_{total}=P_{ite}+P_{hvac}$.


\paragraph{\textbf{Constraints}} The west and east zone temperature is maintained within certain bounds, which is defined as
\begin{equation}
T_{min}\leq T_{west}\leq T_{max},\quad T_{min}\leq T_{east}\leq T_{max}
\end{equation}
where $T_{min}$ and $T_{max}$ is the minimum and maximum temperature.

\paragraph{\textbf{Reward Function}}
We adopt the reward function defined in \cite{rl_gym_env} to train agents using model-free RL approaches e.g. PPO \cite{ppo} and Model Predictive Control. It incorporates total power consumption and temperature violation as defined in Equation~\ref{eq:general_reward_shaping}
where
\begin{align}
r_T=&-\sum_{i=\substack{east\\west}}(e^{(-\lambda_1(T_{i}-T_C)^2)}+\lambda_2([T_{i}-T_{min}]_{+}+[T_{max}-T_{i}]_{+}))\\
r_P=&-P_{total}
\end{align}
This reward function encourages the agents to maintain temperature as close to $T_C$ as possible while minimizing total power consumption. In our experiments, we set $\lambda_1=0.5$, $\lambda_2=0.1$, $\lambda_P=10^{-5}$.

\subsubsection{Building Control Sequences}
\begin{figure}
  \centering
  \includegraphics[width=\linewidth]{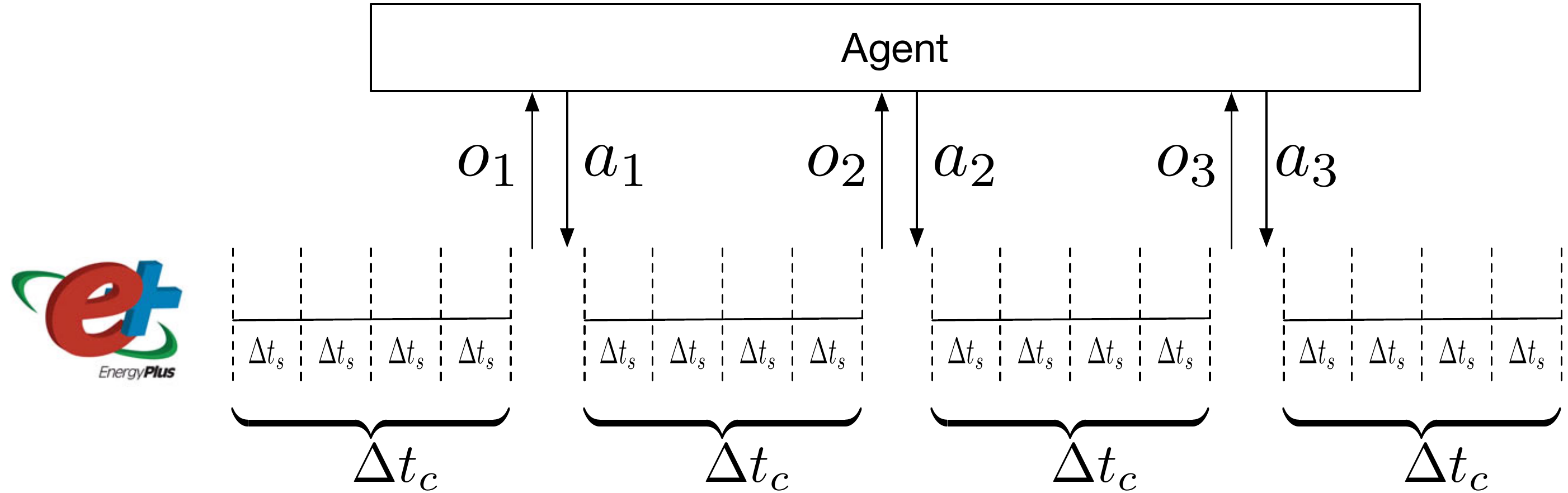}
  \caption{Building Control Sequence}
  \label{fig:control_sequence}
\end{figure}
We show the building control sequences in Figure~\ref{fig:control_sequence}. There are two types of timesteps during the simulation. $\Delta t_s$ refers to the internal simulation timestep in EnergyPlus \cite{energyplus}. Its length varies dynamically ranging from 1 minute/timestep to zone timestep (15 minutes/timestep in this case) to balance simulation precision and speed. $\Delta t_c$ refers to the control interval, which is set to 15 minutes/timestep. During each control interval, the environment sends the \textbf{average} observation of all simulation steps to the agent and the agent send the action back to the simulation environment. The same action is \textbf{repeated} during each simulation step within each control interval.

\subsection{Simulation Setup}
\subsubsection{Parameter Settings}
We show the parameter settings of our model-based RL in Table~\ref{table:parameter_settings}. Notice that the maximum air temperature is set to target temperature since ITE always heat the air.

\begin{table}
  \centering
  \caption{Parameter Settings in Model-based RL}
  \vspace{-1em}
  \begin{tabular}{|c|c|}
    \hline
    Size of historical data & 6240/65 days \\\hline   Batch Size & 128 \\\hline
    Dataset size $|\mathcal{D}|$ & 11520/120 days \\\hline \# of random action sequences $K$ & 8192 \\\hline
    $\Delta TS_{west,, east}$/$\Delta F_{west, east}$  &  $1^\circ$C/1 \\\hline
    $TS_{west,east}$ lower bound/upper bound  &  $13.5^\circ$C/$23.5^\circ$C\\\hline
    $F_{west,east}$ lower bound/upper bound  &  2.5/10.0\\\hline
    Train/Validation split ratio   &  0.8/0.2 \\\hline
    Comfortable Zone  & $23.5^\circ\pm 1.5^\circ$C \\\hline
  \end{tabular}
  \label{table:parameter_settings}
\end{table}

\subsubsection{Data Acquisition}
\paragraph{Weather Data} We use historical weather data bundled with EnergyPlus \cite{energyplus} for our experiments. They are collected and published by the World Meteorological Organization from the following locations:
\begin{itemize}
  \item \textbf{SF:} San Francisco Int’l Airport, CA, USA
  \item \textbf{Golden:} National Renewable Energy Laboratory at Golden, CO, USA
  \item \textbf{Chicago:} Chicago-O’Hare Int’l Airport, IL, USA
  \item \textbf{Sterling:} Sterling-Washington Dulles Int’l Airport, VA, USA
\end{itemize}

\paragraph{IT Equipment Data} The data center server load is simulated by white noise of various amplitude within different time period of a day as shown in Table~\ref{table:ite_simulation_data}.
\begin{table}
  \centering
  \caption{Amplitude of IT Equipment Simulation Data}
  \vspace{-1em}
  \begin{tabular}{|c|c|c|c|c|}
    \hline
    Time Period & 0:00-6:00 & 6:00-8:00 & 8:00-18:00 & 18:00-24:00 \\\hline
    Normalized  & \multirow{2}{*}{0.5}  & \multirow{2}{*}{0.75} & \multirow{2}{*}{1.00} & \multirow{2}{*}{0.80} \\
    Amplitude & & & & \\\hline
  \end{tabular}
  \label{table:ite_simulation_data}
\end{table}


\paragraph{Baseline Controllers} We evaluate the performance of EnergyPlus builtin controllers \cite{energyplus} and Proximal Policy Optimization (PPO) \cite{ppo} based controllers as baselines.

\subsubsection{Neural Network Model Dynamics Architecture}
\begin{figure}
  \centering
  \includegraphics[width=\linewidth]{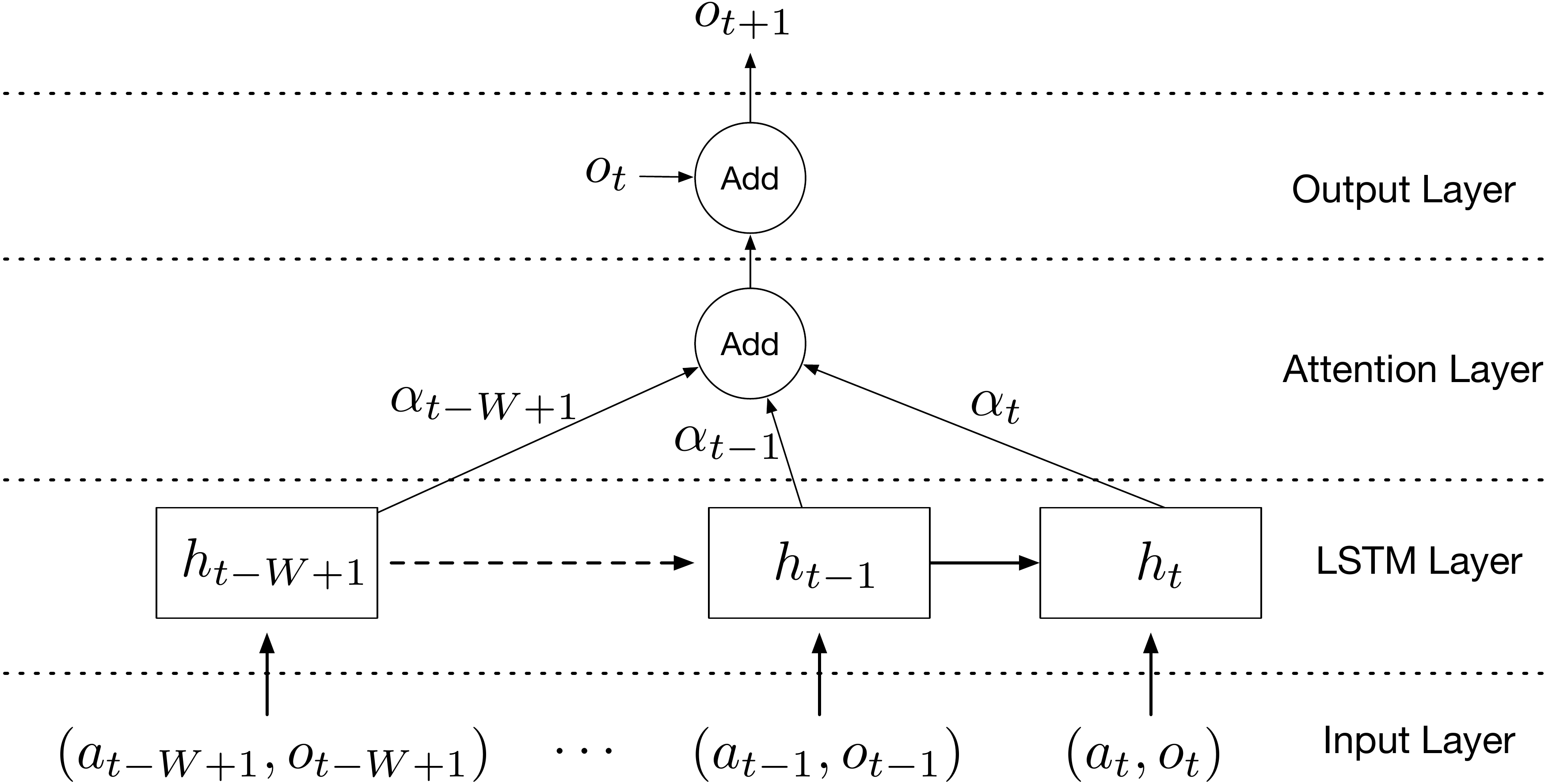}
  \caption{Long short-term memory (LSTM) with attention based system dynamics}
  \label{fig:model_architecture}
\end{figure}
We show the neural network based model dynamics architecture in Figure~\ref{fig:model_architecture}. The idea of the model dynamics architecture is adapted from \cite{attention_relation_extraction}. It contains Input Layer, LSTM Layer, Attention Layer and Output Layer. The Input Layer takes in observations and actions within previous \emph{W} steps. The LSTM Layer is used to extract time series features. The Attention Layer is used to combine those features with automatically learnable weights $\alpha_{t-W+1}, \cdots, \alpha_{t}$. The Output Layer adds $o(t)$ and produce next step observation $o(t+1)$.

\subsection{Performance Metric}
\subsubsection{Learned System Dynamics}
We evaluate the performance of learned system dynamics using \textbf{\emph{H}-step deviation percentage}. Given action sequence $a(t-W+1:t+H)$ and observation sequence $o(t-W+1:t+H)$, we predict $o(t+1:t+H)$ using open-loop prediction as:
\begin{equation}
\begin{split}
\hat{o}(t+h)=&f_{obs,sys}(\hat{o}(t-W+h:t+h),a(t-W+h:t+h))\\
& ,h=1,2,\cdots, H\\
\hat{o}(i)=&o(i), i=t-W+1, \cdots, t
\end{split}
\label{eq:h_step_prediction}
\end{equation}
Then, the \textbf{\emph{H}-step deviation percentage} is given as:
\begin{equation}
d_h=\frac{1}{H}\sum_{h=1}^{H}||\frac{o(t+h)-\hat{o}(t+h)}{o(t+h)}||
\label{eq:h_step_deviation}
\end{equation}

\subsubsection{Controller}
We evaluate \textbf{energy efficiency} of building HVAC controllers by \textbf{Daily Average Power Consumption} and \textbf{temperature requirements} by \textbf{Daily Temperature Violation Rate (TVR)}. The daily average power consumption includes ITE and HVAC power. The daily TVR is defined as the ratio between the number of steps that temperature violates the constraints and the total number of steps within a day.

\subsection{Evaluation of Learned Dynamics}
\begin{figure}
  \centering
  \includegraphics[width=\linewidth]{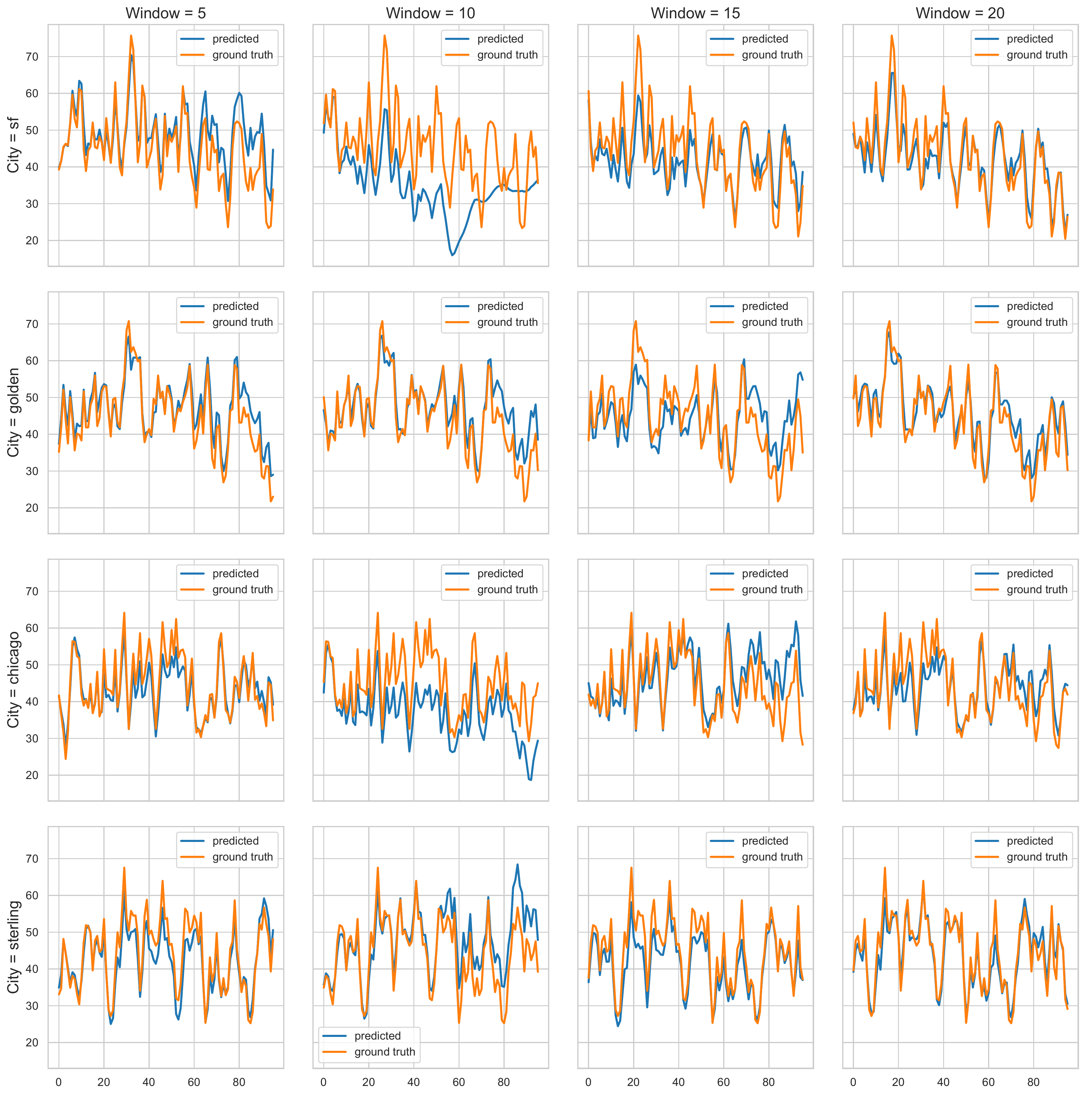}
  \caption{West zone temperature ($^\circ$C) predictions vs. ground truth for random action open-loop predictions}
  \label{fig:west_zone_dynamics}
  \vspace{-1em}
\end{figure}

We collect ground truth observation sequence of the HVAC system by executing random actions sequences. Then, we perform \textbf{\emph{H}-step open loop prediction} defined in Equation~\ref{eq:h_step_prediction} and measure the \textbf{\emph{H}-step deviation percentage} defined in Equation~\ref{eq:h_step_deviation} of various window length and cities in Table~\ref{table:h_step_deviation}. We set \emph{H} to be 96, which amounts to a day. We also show the open-loop predictions vs. ground truth curve for west zone temperature in Figure~\ref{fig:west_zone_dynamics}.

We observe that larger window length yields more accurate predictions in general as shown in Table~\ref{table:h_step_deviation}. In Figure~\ref{fig:west_zone_dynamics}, the west temperature prediction of Chicago with window length 10 even diverges; it may lead to catastrophic outcomes if this predicted temperature is used for optimization. Thus, we set window length to be 20 in all later experiments.


\begin{table}
  \centering
  \caption{H-step Deviation Percentage of Various Window Lengths and Cities. $H=96$}
  \begin{tabular}{c|c|c|c|c|c}
   \multicolumn{2}{c|}{}  & \multicolumn{4}{|c}{Window Length} \\\cline{3-6}
   \multicolumn{2}{c|}{} & 5 & 10 & 15 & 20 \\\hline
   \parbox[t]{2mm}{\multirow{4}{*}{\rotatebox[origin=c]{90}{City}}} & SF & 0.37 & 1.08 &0.19 &0.19 \\\cline{2-6}
   & Golden &0.15 &0.19 &0.29 & 0.14 \\\cline{2-6}
   & Chicago & 0.17 & 0.68 & 0.20 & 0.15 \\\cline{2-6}
   & Sterling & 0.48 & 0.39 & 0.34 & 0.15 \\\hline
  \end{tabular}
  \label{table:h_step_deviation}
\end{table}

\begin{figure}[t!]
  \centering
  \begin{subfigure}[t]{\linewidth}
    \centering
    \includegraphics[width=\linewidth]{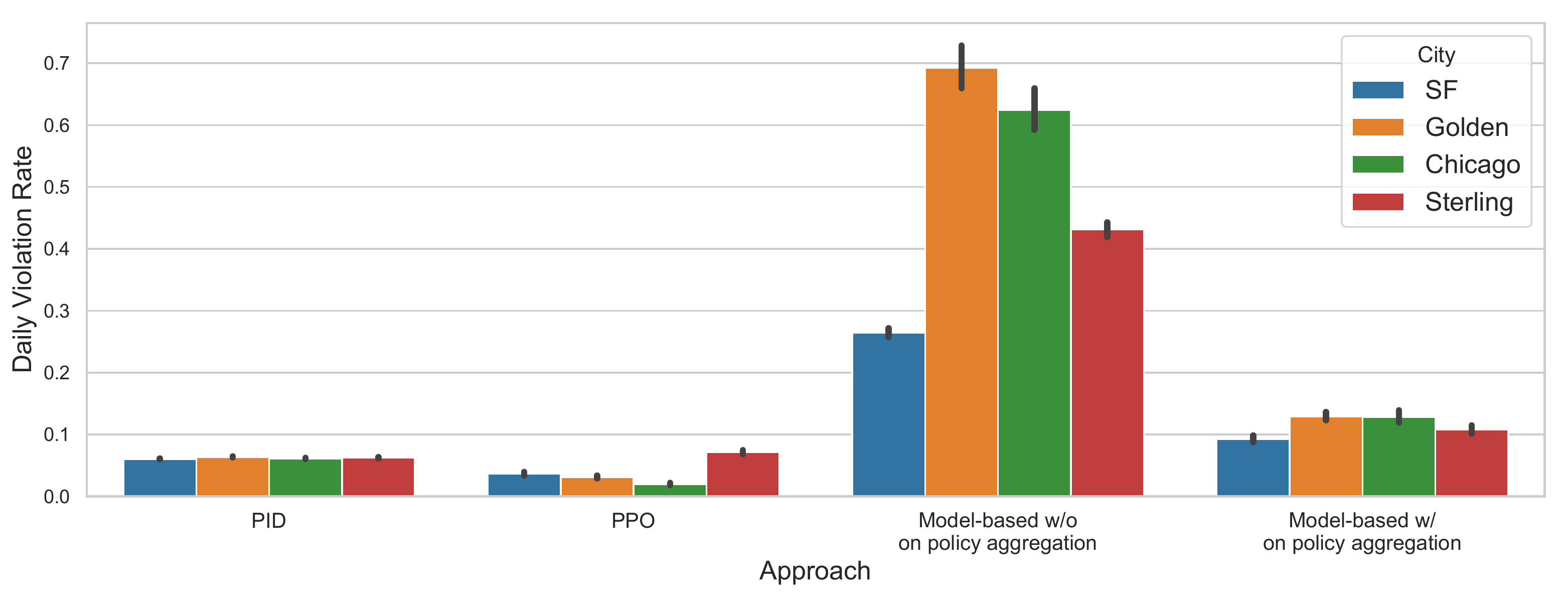}
  \end{subfigure}%
  \vfill
  \begin{subfigure}[t]{\linewidth}
    \centering
    \includegraphics[width=\linewidth]{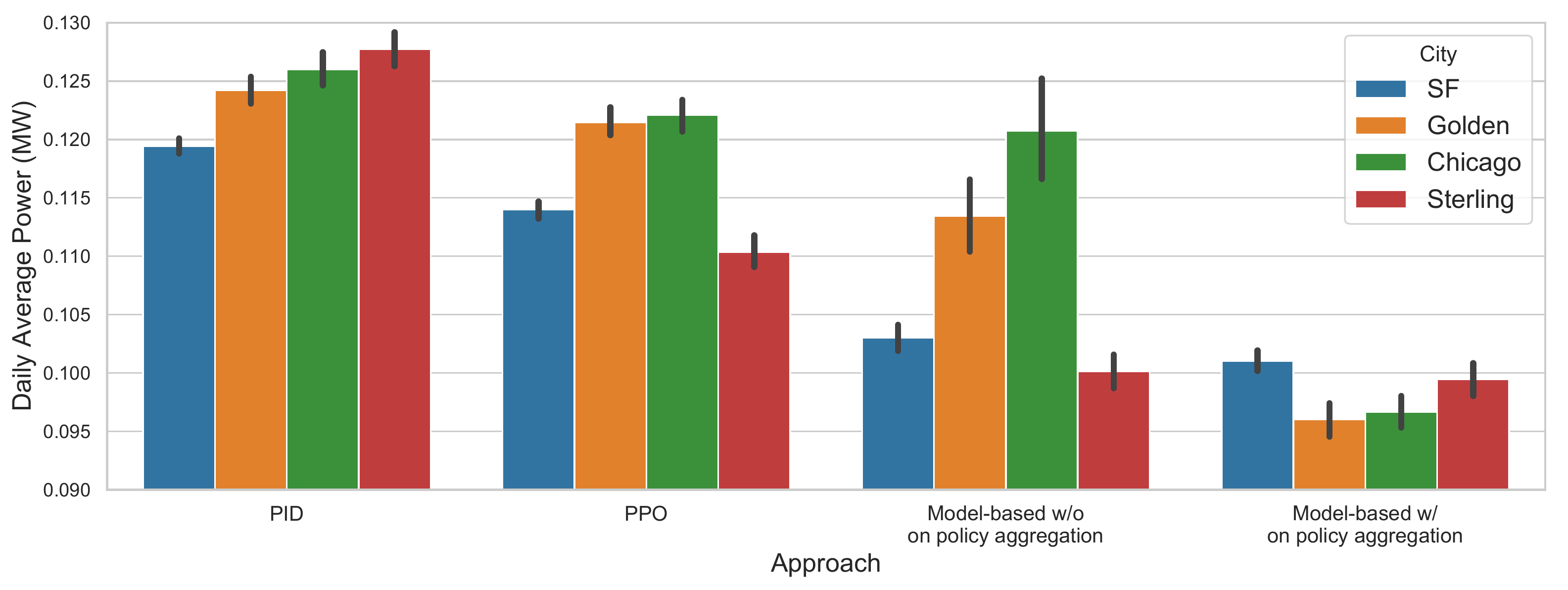}
  \end{subfigure}
  \caption{Daily Violation Rate and Average Power Consumption of various algorithms}
  \label{fig:performance_comparison}
\end{figure}

\begin{figure}
  \centering
  \includegraphics[width=\linewidth]{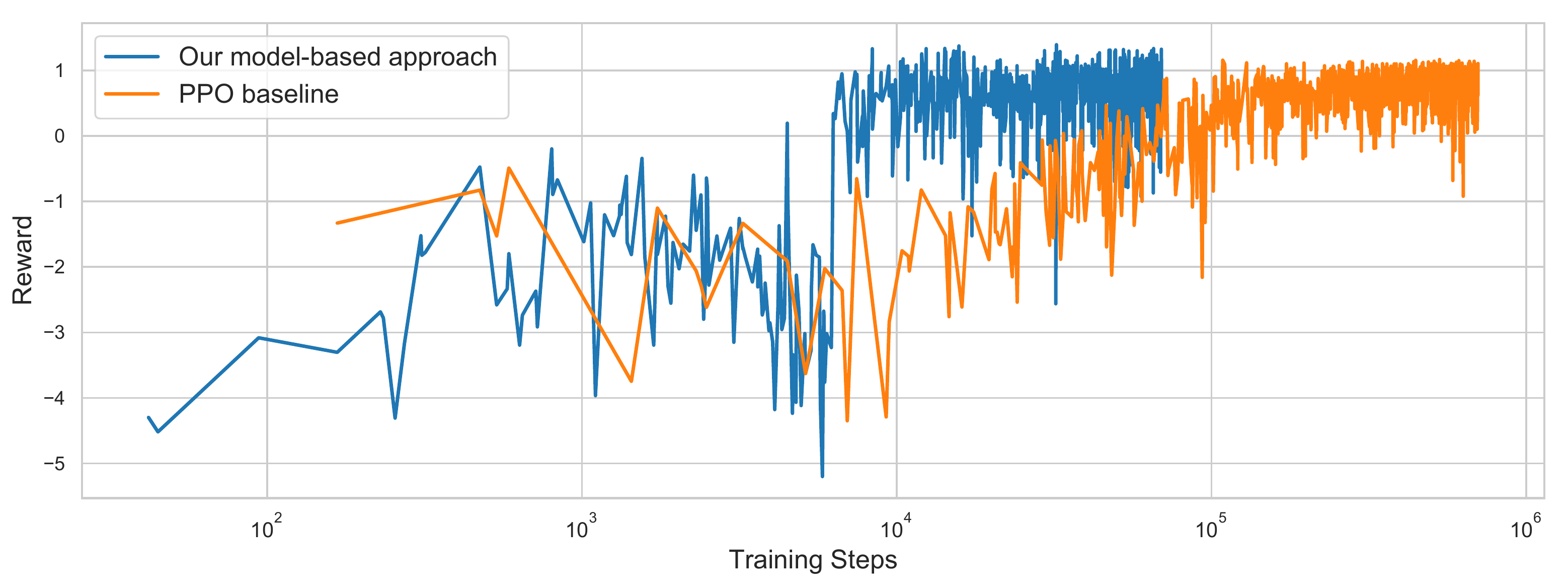}
  \caption{Reward vs. Training steps of model-based and PPO}
  \label{fig:reward_training_steps}
\end{figure}


\subsection{Evaluation of Design Decision}\label{subsection:design_choice}
We perform the evaluation of various design decision by training our MBRL agent in two-room data center environment and report the Daily Average Power Consumption and Daily Temperature Violation Rate.

\begin{figure*}[t!]
  \centering
  \begin{subfigure}[t]{0.67\columnwidth}
    \centering
    \includegraphics[width=\linewidth]{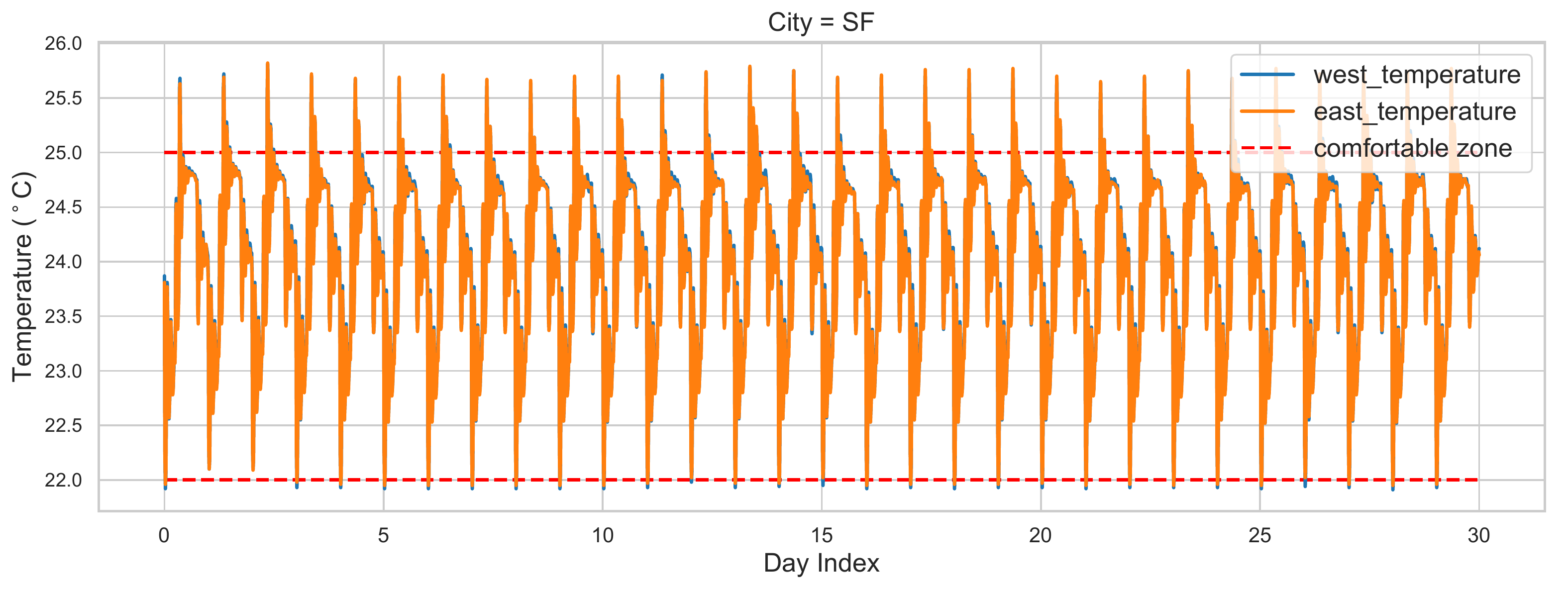}
    \caption{Builtin control}
  \end{subfigure}
  \hfill
  \begin{subfigure}[t]{0.67\columnwidth}
    \centering
    \includegraphics[width=\linewidth]{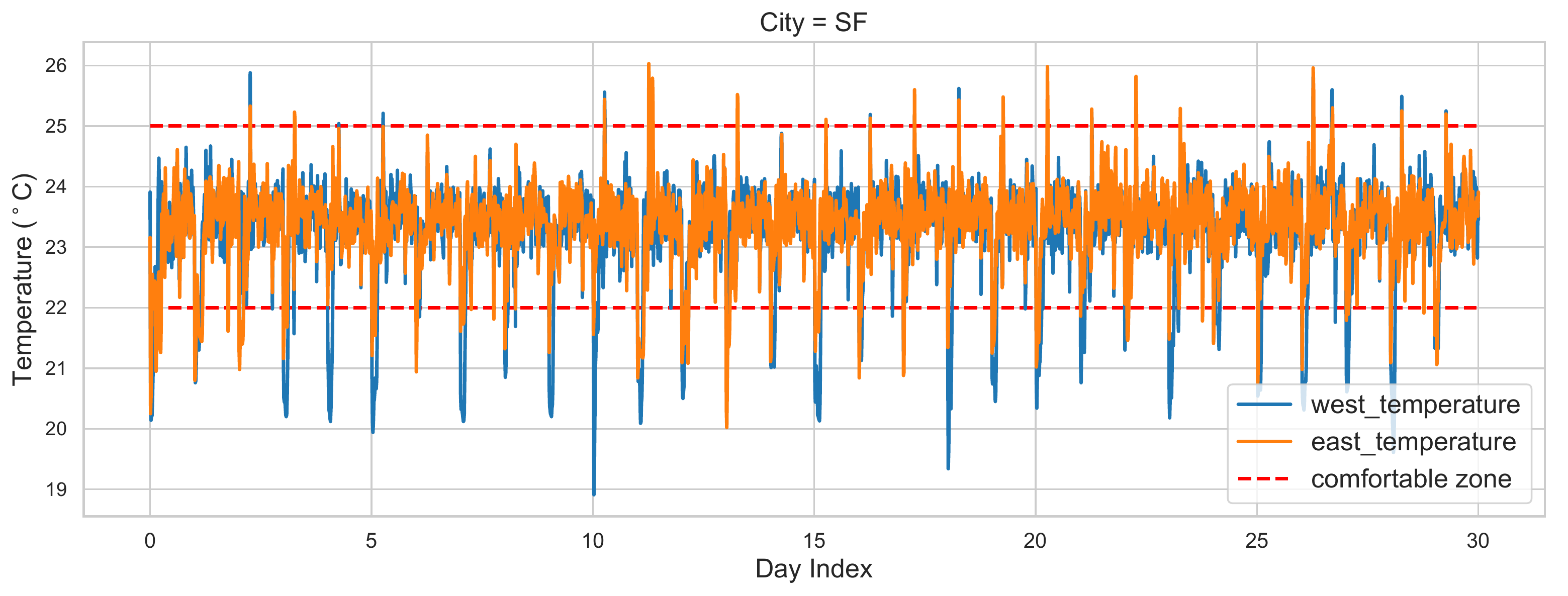}
    \caption{Model-based Reinforcement Learning with Neural Network Dynamics}
  \end{subfigure}
  \hfill
  \begin{subfigure}[t]{0.67\columnwidth}
    \centering
    \includegraphics[width=\linewidth]{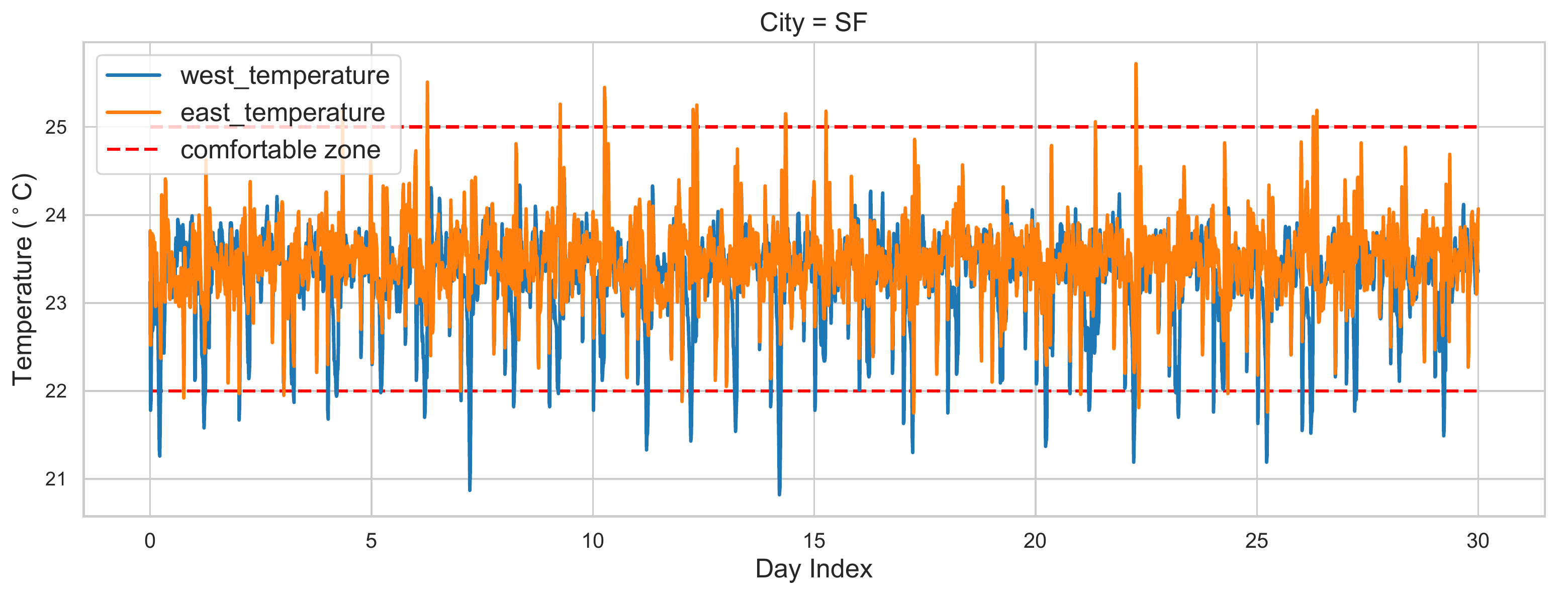}
    \caption{Model-free Reinforcement Learning (PPO)}
  \end{subfigure}
  \caption{One month test temperature curve of various algorithms}
  \label{fig:temperature_curve}
\end{figure*}

\subsubsection{Training Epochs}
\begin{figure}[!t]
  \centering
  \begin{subfigure}[h]{\linewidth}
    \centering
    \includegraphics[width=\linewidth]{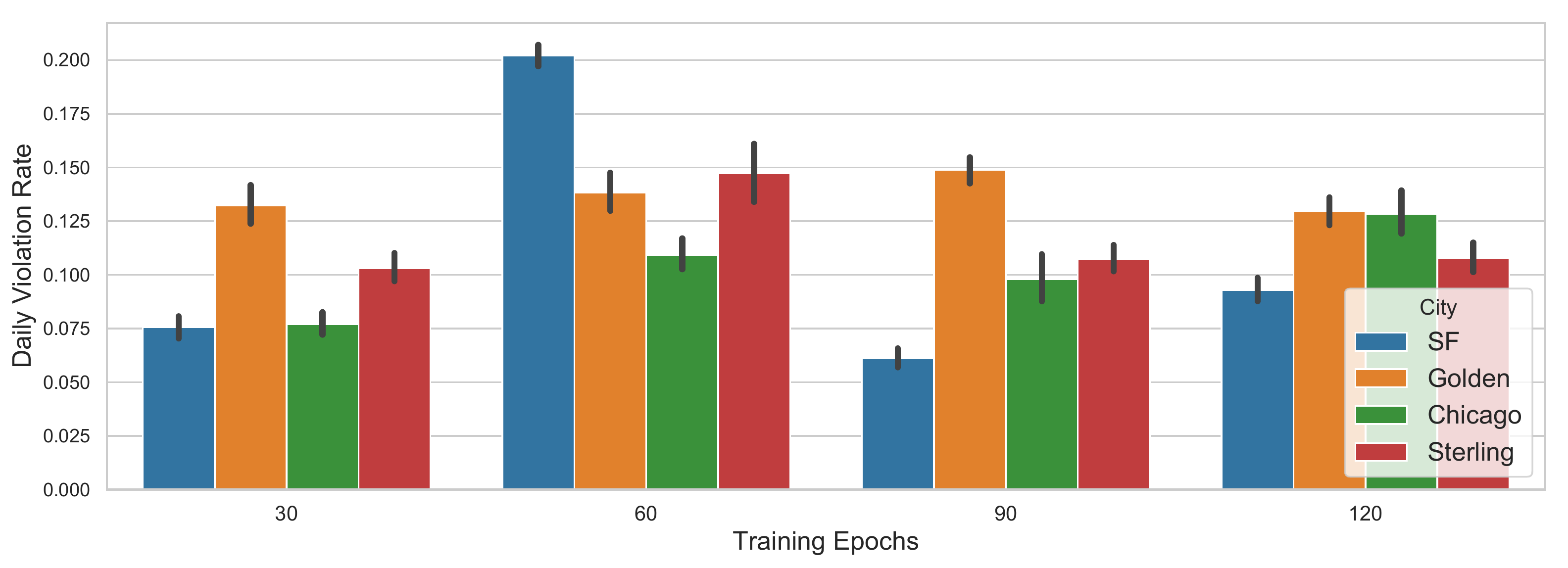}
  \end{subfigure}%
  \vfill
  \begin{subfigure}[h]{\linewidth}
    \centering
    \includegraphics[width=\linewidth]{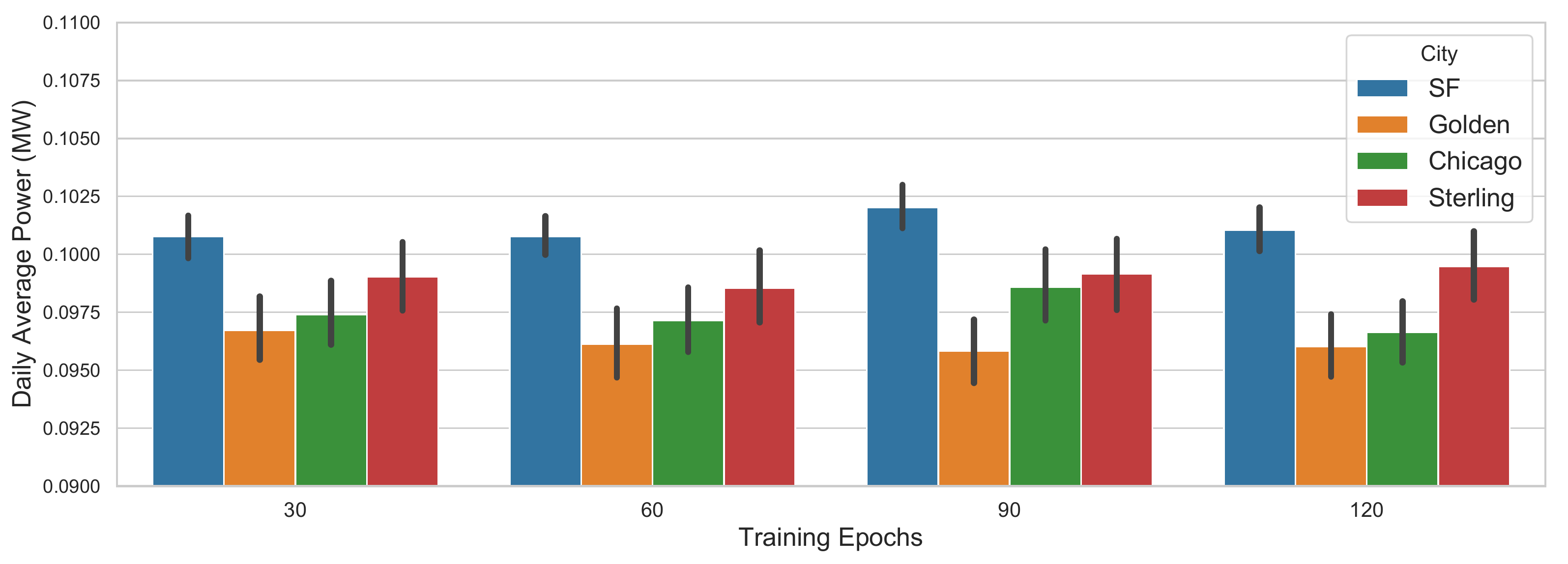}
  \end{subfigure}
  \caption{Daily Violation Rate and Average Power Consumption of various dynamics model training epochs}
  \label{fig:training_epoch_performance}
\end{figure}

It refers to the number of epochs $M$ to fit model dynamics using current dataset in Algorithm~\ref{algo:model_rl_hvac_control} and Algorithm~\ref{algo:model_rl_hvac_control_imitation}. The model may be underfit and fail to predict future observations accurately if $M$ is too small. On the contrary, the model may overfit to the current dataset and hard to adapt when the data distribution changes over time if $M$ is too large. We vary training epochs by 30, 60, 90 and 120 and show the control performance in Figure~\ref{fig:training_epoch_performance}. It is shown that the performance of training epochs 30 works best.

\subsubsection{MPC Horizon}
\begin{figure}[t!]
  \centering
  \begin{subfigure}[t]{\linewidth}
    \centering
    \includegraphics[width=\linewidth]{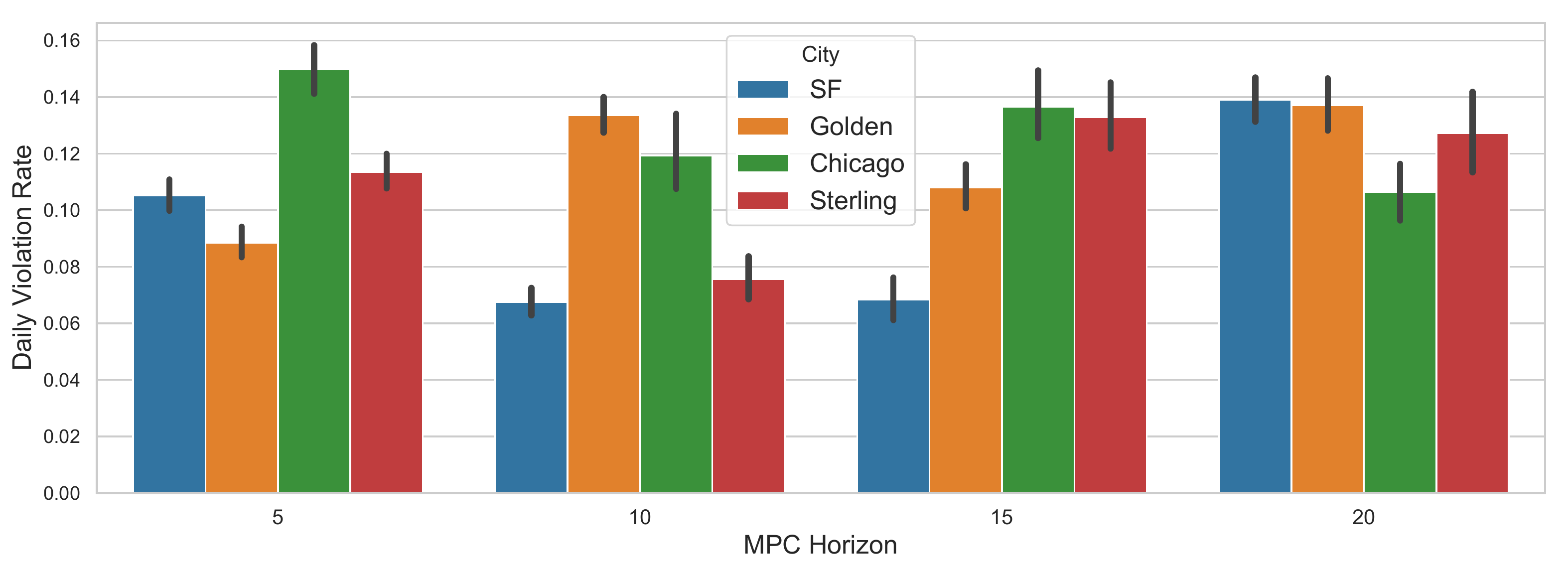}
  \end{subfigure}%
  \vfill
  \begin{subfigure}[t]{\linewidth}
    \centering
    \includegraphics[width=\linewidth]{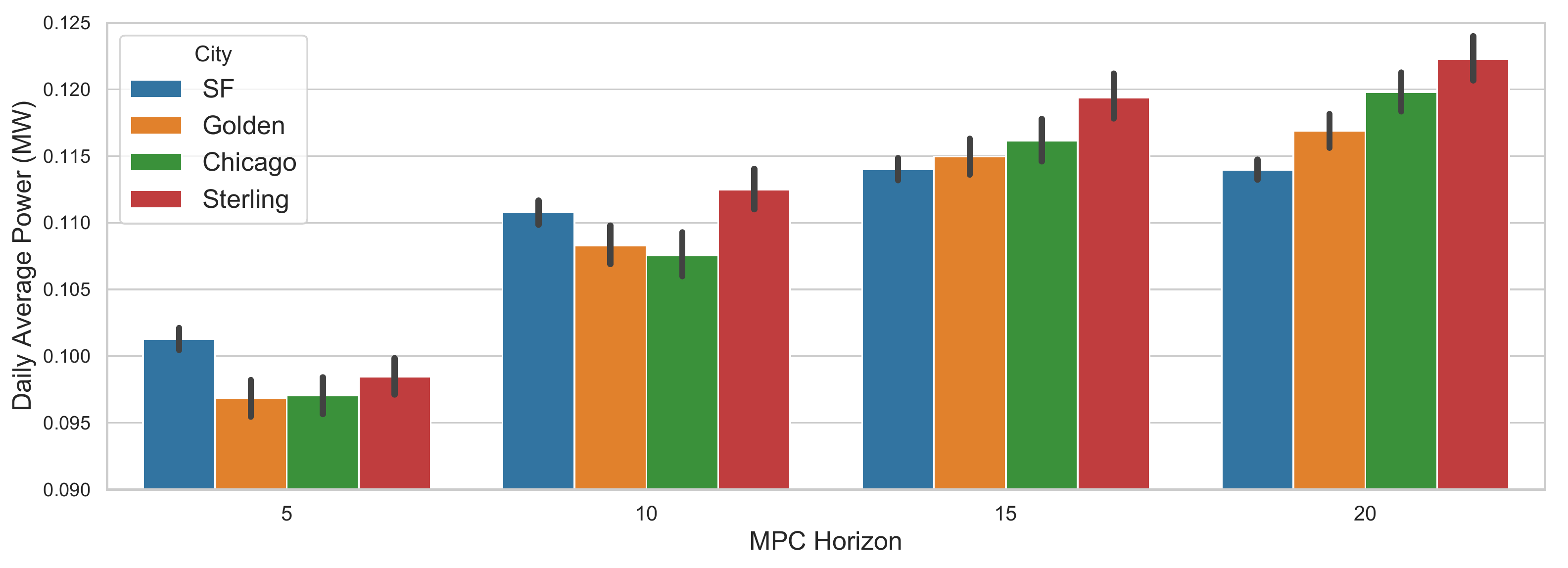}
  \end{subfigure}
  \caption{Daily Violation Rate and Average Power Consumption of various MPC horizons}
  \label{fig:mpc_horizon_performance}
  \vspace{-1em}
\end{figure}

The MPC horizon refers to the number of steps to look ahead when performing model predictive control (MPC). Small MPC horizon results in more greedy actions that may fail to overcome the inertia of thermal dynamics. Large MPC horizon may produce worse actions since the prediction errors aggregate as the horizon becomes larger. We vary the MPC horizons by 5, 10, 15 and 20 and show the control performance in Figure~\ref{fig:mpc_horizon_performance}. Results show that the controller works best when MPC horizon equals 5 steps.

\subsubsection{On Policy Frequency}
\begin{figure}[t!]
  \centering
  \begin{subfigure}[t]{\linewidth}
    \centering
    \includegraphics[width=\linewidth]{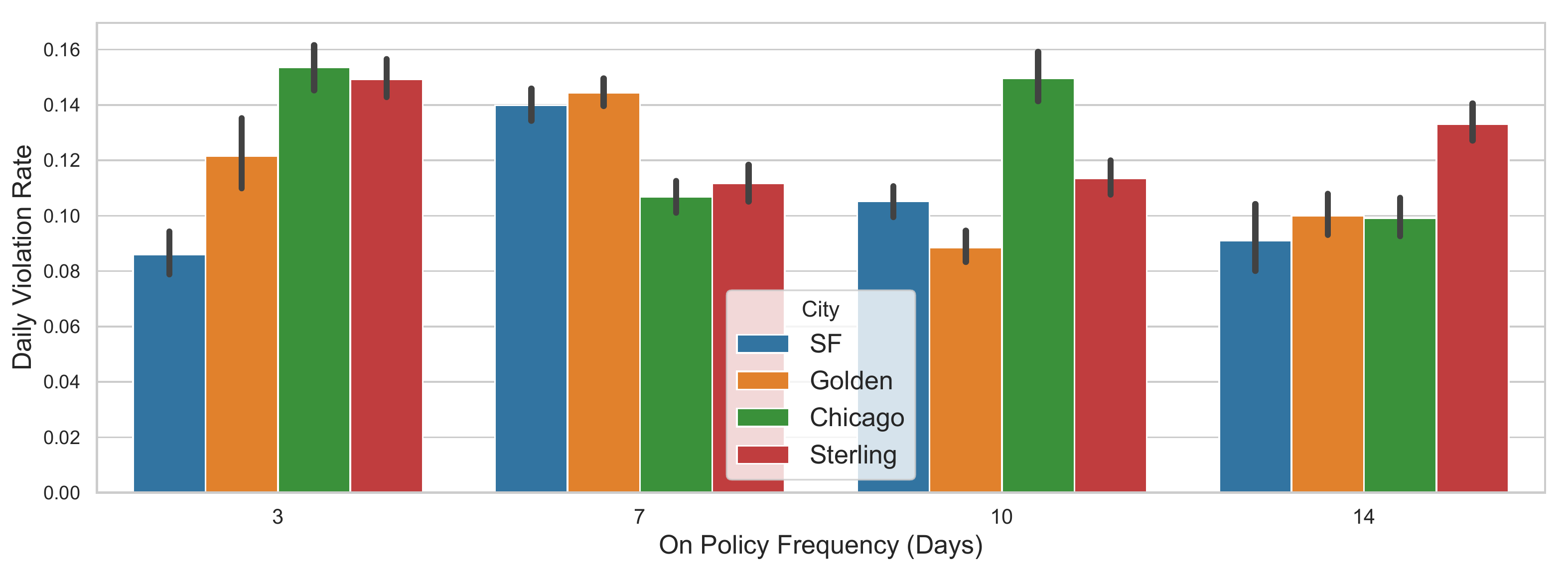}
  \end{subfigure}%
  \vfill
  \begin{subfigure}[t]{\linewidth}
    \centering
    \includegraphics[width=\linewidth]{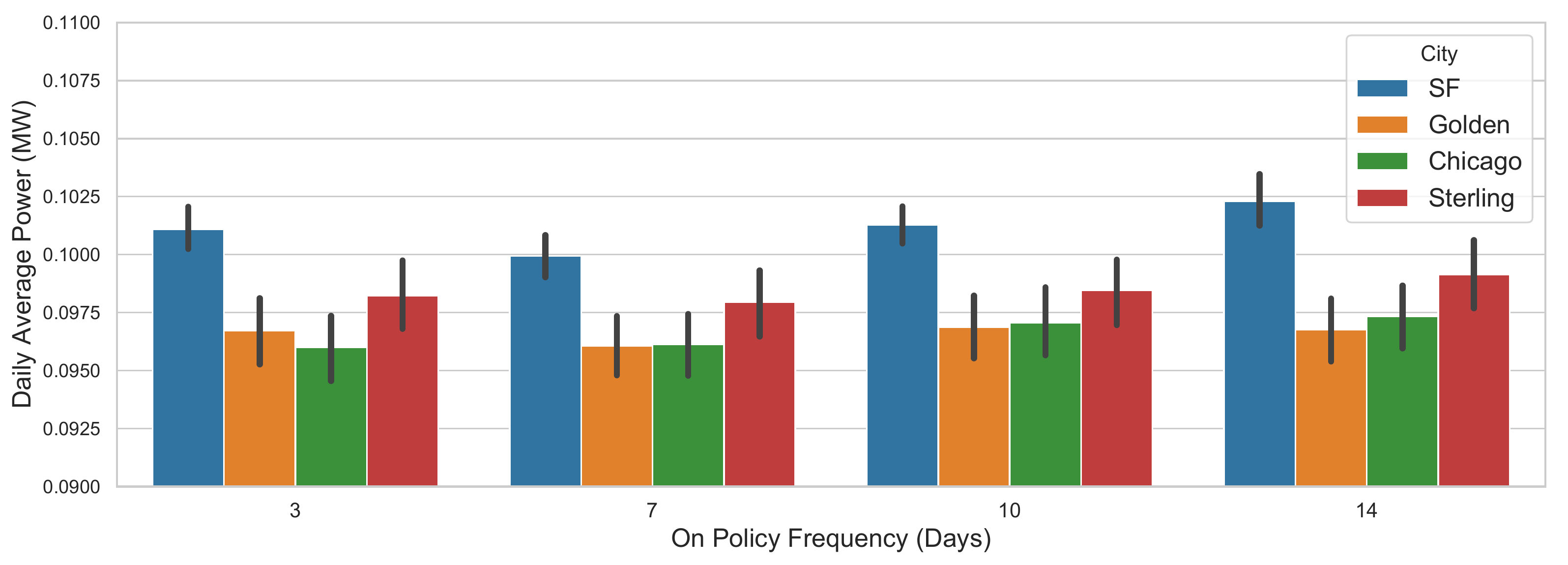}
  \end{subfigure}
  \caption{Daily Violation Rate and Average Power Consumption of various on policy steps}
  \label{fig:on_policy_frequency_performance}
\end{figure}

The on policy frequency refers to the number of days the agent uses the current system dynamics to perform model predictive control. It refers to $T$ in Algorithm~\ref{algo:model_rl_hvac_control} and Algorithm~\ref{algo:model_rl_hvac_control_imitation}. Smaller $T$ results in more system dynamics training iterations with the risk of overfitting to the current data distribution. Large $T$ results in fewer system dynamics training iterations with the risk of failing to adapt to the new data distribution. We vary the on policy frequency by 3, 7, 10 and 14 days and show the control performance in Figure~\ref{fig:on_policy_frequency_performance}. Results show that the controller works best when on policy frequency equals 7 days.

\subsubsection{Imitation Learning}
\begin{figure}[t!]
  \centering
  \begin{subfigure}[t]{\linewidth}
    \centering
    \includegraphics[width=\linewidth]{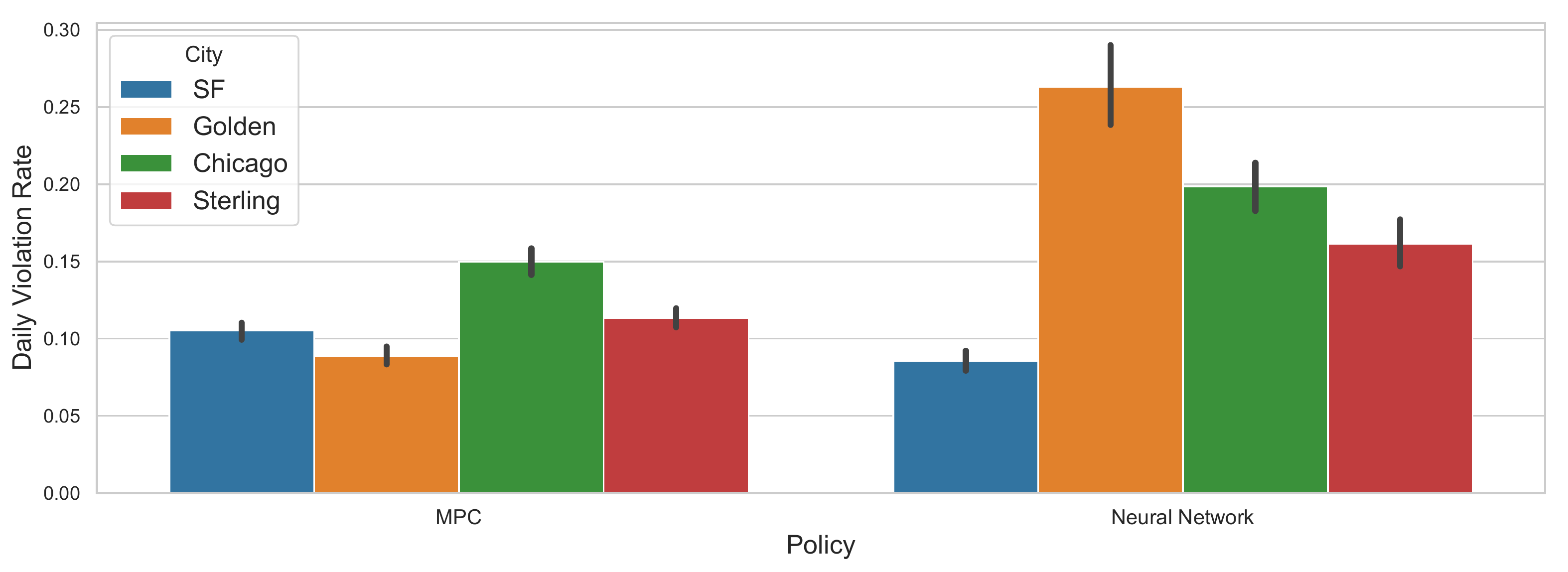}
  \end{subfigure}%
  \vfill
  \begin{subfigure}[t]{\linewidth}
    \centering
    \includegraphics[width=\linewidth]{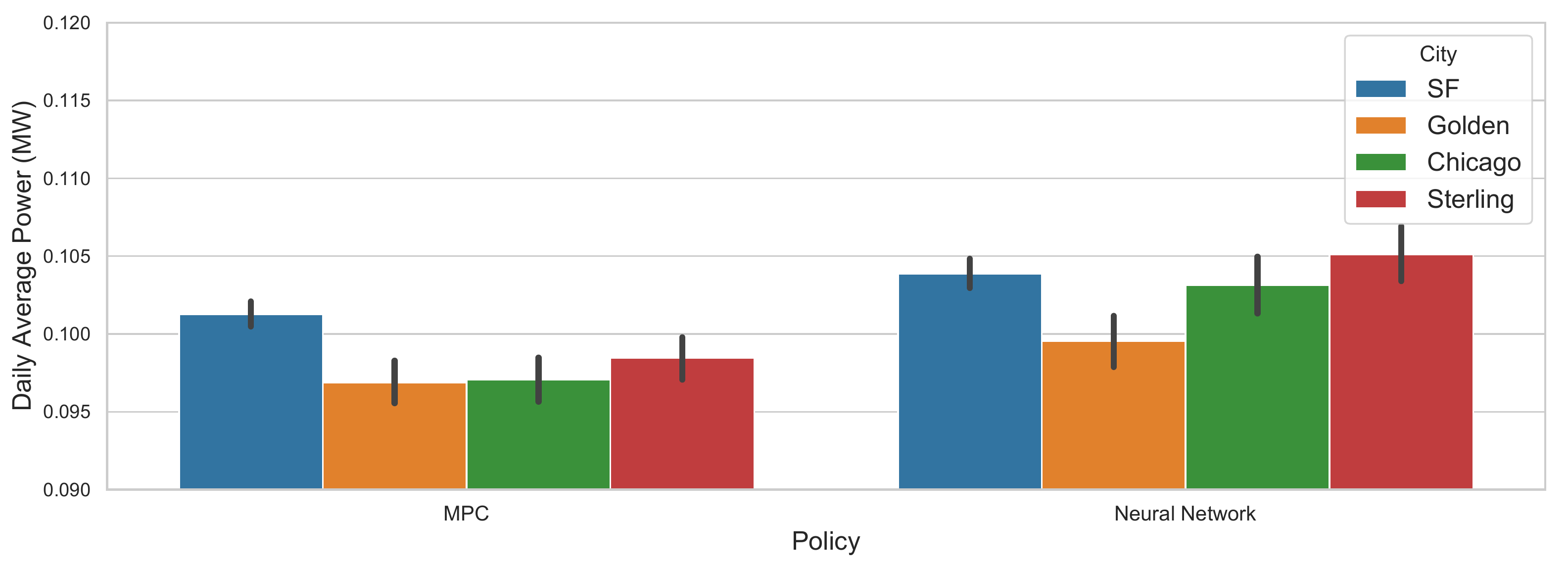}
  \end{subfigure}
  \caption{Daily Violation Rate and Average Power Consumption of MPC vs. Imitation Learning}
  \label{fig:imitation_learning_performance}
\end{figure}

The main issue of model predictive control is the slow processing speed that fails for real-time control with finer control intervals. Thus, we train a neural network policy that mimics the output of MPC for real time control. We show the control performance of both MPC and neural network policies in Figure~\ref{fig:imitation_learning_performance}. We observe that the neural network policy works worse because the historical data collected from MPC may not cover all future control scenarios and the neural network fail to generalize.

\subsection{Model Predictive Control using Learned Dynamics on Benchmarks}
We evaluate our learned model-based controller using the best configuration found in Section~\ref{subsection:design_choice}. As comparison, we evaluate the performance of model-free approach (PPO) and builtin controller on the same environment.

\subsubsection{Effectiveness of our approach in satisfying temperature requirements}

We show one month controlled temperature curve of various algorithms in Figure~\ref{fig:temperature_curve}. We observe that both model-based and model-free RL manages to maintain the west and east zone temperature around $23.5^\circ $C.

\subsubsection{Performance Comparison}
We show the daily violation rate and daily average power in Figure~\ref{fig:performance_comparison}. Compared with model-free approach, our model-based approach achieves $17.1\% \sim 21.8\%$ power reduction with slightly increased violation rate. Also, the on-policy data aggregation plays critical role in addressing observation distribution shifting as shown by the violation reduction.

\subsubsection{Computation Speed Analysis}
As the experiments suggest, the MPC takes more than 30 seconds to finish while the neural network policy takes less than 1 second for inference. It indicates the advantages of neural network policy in high frequency control. However, possible failure modes need to be prevented by proper interference due to the weak robustness of neural network policy.

\subsubsection{Convergence Analysis}
Although, model-free RL approach achieves similar control performance with model-based approach, it requires tremendous amount of trial-and-error with the actual environments, which is not possible in real systems. We show the reward vs. training steps of model-based approach and PPO in Figure~\ref{fig:reward_training_steps}. We notice that model-free approaches requires approximately 10x more training steps to converge to the same performance level as model-based approach.
\section{Conclusion and Future Work}
In this paper, we propose a model-based reinforcement learning approach for building HVAC scheduling via neural network based model approximation. 
We first learn the system dynamics using neural network by collecting data through interactions with the system. 
Then, we use the learned system dynamics to perform model predictive control using random-sampling shooting method.
To overcome system distribution shift such as outside temperature and IT equipment load schedules, we retrain the dynamics using on-policy data aggregation. Experiments show that our approach achieves significant improvement of power reduction compared with baseline controllers. Compared with model-free reinforcement learning approach (PPO), our approach improves the sample efficiency by 10x.

Future work includes conducting experiments with more complex systems containing larger observation and action space, or even probabilistic system dynamics. It is also useful to experiment with time-variant system objective and constraints and see how our model-based approach advantages over model-free approaches.


\bibliographystyle{bib/ACM-Reference-Format}
\bibliography{bib/buildsys19_chi}

\end{document}